\documentclass[10pt,letterpaper]{article}
\usepackage{opex3}
\begin{document}
\title{The Role of Nuclear Coulomb Attraction in Nonsequential Double Ionization of Argon Atom}
\author{Qing Liao, Yueming Zhou, and Peixiang Lu$^{ \dagger}$}
\address{Wuhan National Laboratory for Optoelectronics, Huazhong University of
Science and Technology, Wuhan 430074, P. R. China}
\email{$^\dagger$Corresponding author: lupeixiang@mail.hust.edu.cn}

\begin{abstract}
The role of nucleus in strong-field nonsequential double ionization
of Ar atoms is investigated using three-dimensional classical
ensembles. By adjusting the nuclear Coulomb potential, we can
excellently reproduce the experimental correlated electron and ion
momentum spectra with laser intensities above the recollision
threshold [Phys. Rev. Lett. 93, 263001 (2004)] and below the
recollision threshold [Phys. Rev. Lett. 101, 053001 (2008)]
quantitatively. Analysis reveals the detailed electronic dynamics
when the nuclear Coulomb attraction plays a key role in the
recollision process of nonsequential double ionization of Ar atoms.
Comparison between our results for Ar and those for He shows that
atom species have a strong influence on nonsequential double
ionization.
\end{abstract}
\ocis{(020.4180) Multiphoton processes; (260,3230) Ionization;
(270.6620) Strong-field processes.quadrants.}

\section{Introduction}

Nonsequential double ionization (NSDI) of atoms and molecules has
provided a profound understanding of laser-mater interaction and
electron correlation [1-7]. In order to explore the physical
mechanism of NSDI, a series of experimental  and theoretical studies
[2-6, 8-15] have been performed. The rescattering model [16] is now
widely accepted to be the basic mechanism of NSDI at near infrared
(NIR) wavelengths. In the rescattering picture, one electron firstly
escapes from the atom near the peak of the electric field. After
changing its direction, the laser field drives the electron back to
the parent ion and then causes the ionization of the bound electron
through the inelastic recollision. According to this model, if the
second electron is ionized directly through the recollision, the
process is called (e, 2e) ionization, i.e. recollision-ionization
(RCI), whereas if the second electron is excited through the
recollision with subsequent field-ionization by the laser field, the
process is called recollision excitation with subsequent ionization
(RESI) [17]. Based on this inelastic rescattering mechanism, at low
laser intensities, RESI mechanism dominates NSDI and the
longitudinal momentum distributions of the doubly charged ions
exhibits a single peak structure near zero momentum [3]. While at
high laser intensities, RCI mechanism dominates NSDI and the
longitudinal momentum distributions of the doubly charged ions will
exhibit a double-peak structure at nonzero momenta [5, 17].

It has been shown that the microscopic dynamics of NSDI is dependent
on the atom species because of the different atom structures
\cite{15, 17}. Recent experimental studies [18, 19] showed that the
nuclear Coulomb attraction plays a significant role in forming the
finger-like pattern of the correlated electron momentum
distributions from strong-field double ionization of helium. The
important role of nuclear Coulomb attraction was firstly
demonstrated in \cite{20} by the three-dimensional (3D) classical
ensembles model \cite{13}. In order to quantitatively reproduce the
experimental results of NSDI for different atoms and predict new
experimental phenomena, employing the proper nuclear Coulomb
potentials for different atoms is of vital importance.

In this paper, we report the calculation results of correlated
momentum spectra from NSDI of Ar atoms by intense laser pulses with
intensities above and below the recollision threshold using the
three-dimensional (3D) classical ensembles [13, 20]. In this model,
the nuclear Coulomb potential is described by soft-core Coulomb
potential. By adjusting the nuclear Coulomb potential to a proper
form, we can excellently reproduce two experimental results
\cite{5,10}. Back analysis reveals that the nuclear Coulomb
attraction has a strong influence on the microscopic dynamics in the
recollision process. By comparing the nuclear Coulomb potential of
He atoms, we can infer that the inner-shell electrons also have a
significant influence on NSDI of Ar atoms.

\section{The classical Ensemble Model}

The 3D classical ensembles has been described in detail in \cite{13,
20}. This model has achieved great successes in studying the NSDI
process \cite{13, 14, 20, 21}. This classical model discards all
aspects of quantum mechanics. In this model, the first ionization
occcurs over a laser-suppressed barrier \cite{20}, not through
tunneling, which is retained within a semi-classical model
\cite{Chen}. However, the classical model is adequate to generate
very strong two-electron correlation in NSDI \cite{Ho}. Many
recollision details \cite{13,14,20,Ho,24} are revealed by this
model. By back-tracing the trajectories, recollision mechanisms can
be classified into RCI mechanism if the delay time between closest
recollision and double ionization is less than 0.25 laser cycle and
RESI mechansim if larger than 0.25 laser cycle \cite{24}. This
classification of recollision mechanisms is consistent with that in
\cite{17}. RCI trajectories give distributions of doubly charged
ions peaked at nonzero momenta and RESI trajectories give a
distribution of ions peaked at zero momentum. Another classification
of trajectories: nonzero and zero momenta trajectories in \cite{Ho},
are in accordance with RCI and RESI trajectories respectively. In
our simulations, a large ensemble containing one or two millions
classical electron pairs is used. The evolution of the system is
determined by the classical equation of motion (Atomic units are
used throughout the paper unless otherwise stated):
$d^2\vec{r_{i}}/dt^2=-\vec{E}(t)-\vec{\nabla}[V_{ne}(\vec{r}_{i})+V_{ee}(\vec{r}_{1},\vec{r}_{2})]$,
where the subscript $i$ is denotes the two different electrons and
$\vec{E}(t)$ is the linearly polarized electric field. The
nucleus-electron and the electron-electron interaction are
represented by 3D soft-Coulomb potential
$V_{ne}=-2/\sqrt{r^2_i+a^2}$ and
$V_{ee}=1/\sqrt{(\vec{r}_1-\vec{r}_2)^2+b^2}$, respectively. To
obtain the initial value, the ensemble is populated starting from a
classically allowed position for the argon ground-state energy of
-1.59 a.u. The available kinetic energy is distributed between the
two electrons randomly in momentum space. Each electron is given a
radial velocity only, with sign randomly selected [13]. Then the
electrons are allowed to evolve a sufficient long time(100 a.u.) in
the absence of the laser field to obtain stable position and
momentum distributions. To avoid autoionization and ensure stability
in three dimensions, the nuclear soft-core parameter $a$ is set to
be larger than 1.26 a.u. In this paper, we set to $a=1.5$ a.u. Then,
in order to explore the influence of nuclear Coulomb potential on
the process of recollision, we can decrease the value of $a$ after
first ionization. To conserve energy we offset the decrease in the
potential energy of each electron with a kinetic boost for its
radial motion [20]. The soft-core parameter $b$ is set to be 0.05
a.u. throughout the whole process. The nucleus remains fixed at the
origin at all times. After the laser pulse is turned off, if both
electrons have positive energy, we define double ionization.

\section{Results and Discussions}
Firstly, we show the calculation results of NSDI of Ar atoms by
intense laser pulses with an intensity well above the recollision
threshold. The laser parameters are employed as the same as in the
experiment of Ref.\cite{5}. The laser pulse is a linear polarized
few-cycle pulse. Its electric field is
$\vec{E}(t)=\vec{e}{_x}E_0\sin^2({\pi}t/Nt)\cos[{\omega}(t-NT/2)+{\phi}]$,
where $\vec{e}{_x}$ is the polarization vector. ${E{_0}}$, $\omega$,
${T}$ and $\phi$ are the amplitude, frequency, period and
carrier-envelope phase (CEP), respectively. The wavelength
$\lambda$=760 nm, the carrier frequency $\omega$=0.06 a.u., the
intensity $I$=3.5$\times$$10^{14}$ W/cm$^2$ and period of laser
cycle $T$=2$\pi$/$\omega$. The pulse contains five laser cycles,
where the full-width at half-maximum is about 5 fs.

\begin{figure}[htb]
\centering\includegraphics[width=6cm]{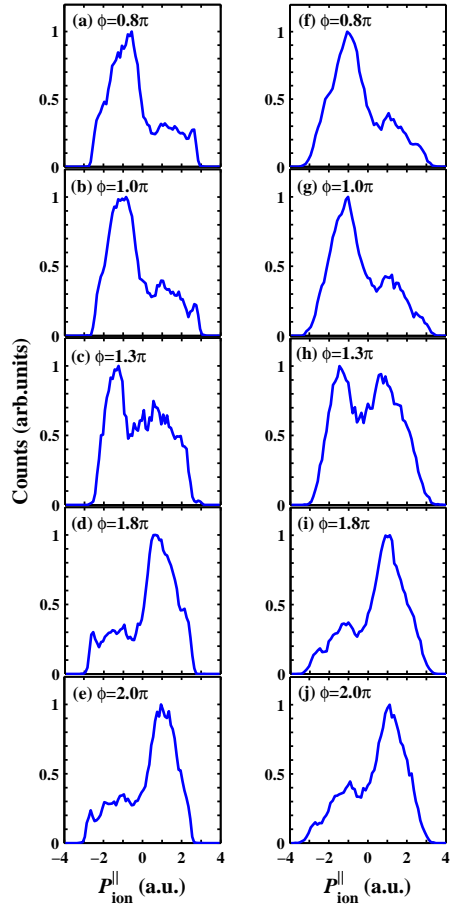} \caption{ (color
online). Momentum spectra of the Ar$^{2+}$ ions for different CEPs
$\phi=0.8\pi$ (a, f), $1.0\pi$ (b, g), $1.3\pi$ (c, h), $1.8\pi$ (d,
i) and $2.0\pi$ (e, j), respectively. According to the laser
parameters, $\pm$4$\sqrt{U_p}$=$\pm$3.33 a.u. The soft-core
parameters $a$=1.5 a.u. for Figs. 1(a)-1(e) and $a=1.0$ a.u. for
Figs. 1(f)-1(j) after first ionization. }
\end{figure}

The recoil momentum of the Ar$^{2+}$ ion is the sum of the two
electron momenta with a reversed direction [22], i.e.
$P^{||}_{ion}=-(P^{||}_{e1}+P^{||}_{e2})$, because the photon
momentum is negligible. The final-state longitudinal momentum
spectra of the doubly charged ions as a function of the CEP is shown
in Fig. 1, where the values of the CEP $\phi=0.8\pi$(a, f),
$1.0\pi$(b, g), $1.3\pi$(c, h), $1.8\pi$(d, i) and $2.0\pi$(e, j),
respectively. In Figs. 1(a)-1(e), the soft-core parameter a keeps
unchanged (1.5 a.u.) in the whole double ionization process. Though
the ion momentum distributions are sensitive to the CEP of the
pulse, they are different from the experimental results (Figs.
2(a)-2(e) of Ref. \cite{5}). When decreasing the soft-core parameter
a after first ionization to enhance the role of nuclear Coulomb
attraction in recollision process, we find that the calculation
results are in good agreement with the experimental results only for
$a=1$ a.u., as shown in Figs. 1(f)-1(j). The ion longitudinal
momentum spectra are prominent asymmetric. A change for the
asymmetry appears with the center of gravity of the ion momentum
distributions shifting from negative to positive as the CEP
increases from $0.8\pi$ to $2\pi$. While the CEP $\phi=1.3\pi$, a
double-hump structure with maxima of nearly equal yields appears. As
the CEP $\phi=0.8\pi$ (or $1.8\pi$), the asymmetry reaches the
maximum. Compared with Figs. 1(a)-1(e), the ion momentum spectra of
Figs. 1(f)-1(j) exhibit three prominent changes. Firstly, the range
of the momentum spectra becomes broader and equal to the range of
[$-4\sqrt{U_p}$, $4\sqrt{U_p}$], where ${U_p=E_0^2/(4\omega^2)}$ is
the ponderomotive energy and $4\sqrt{U_p}=3.33$ a.u. Secondly, the
ion yield around zero momentum increases, especially for Fig. 1(h).
In addition, the double-hump structure in Fig. 1(h) is more
symmetric than that in Fig. 1 (c). All these features in Figs.
1(f)-1(j) are quantitatively consistent with the experimental
results of Ref. \cite{5}.

\begin{figure}[htb]
\centering\includegraphics[width=7cm]{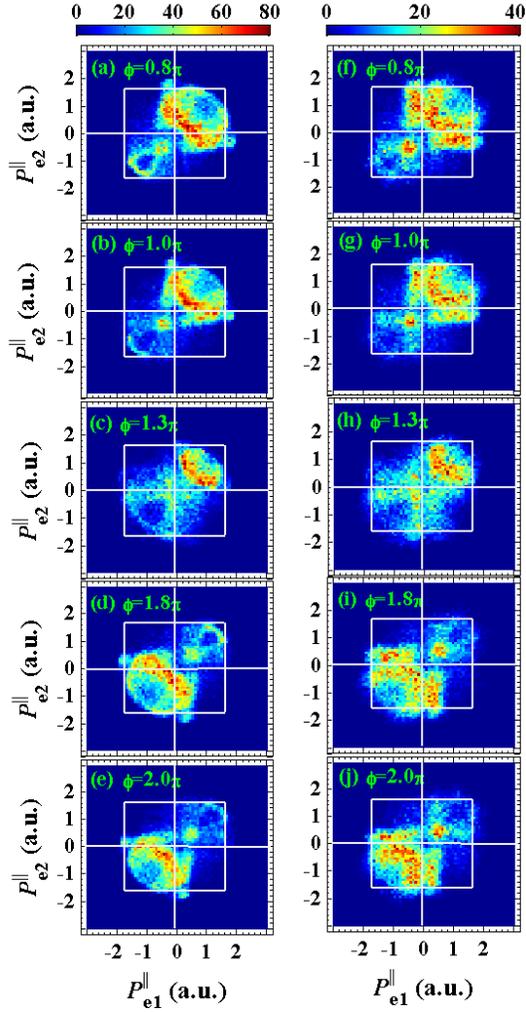} \caption{ (color
online). Correlated electron momentum distributions parallel to the
laser polarization direction of argon NSDI for different CEPs. The
laser parameters are the same as Fig. 1. The soft-core parameters
$a$ =1.5 a.u. for Figs. 2(a)-2(e) and 1.0 a.u. for Figs. 2(f)-2(j)
after first ionization. White boxes indicate $\pm$2$\sqrt{U_p}$
=$\pm$1.66 a.u.}
\end{figure}

The strong phase dependence of the differential momentum spectra of
Ar$^{2+}$ ions from NSDI by linearly polarized few-cycle pulses is
explained in Ref. \cite{5} using a classical model and Ref.
\cite{23} by solving the one-dimensional time-dependent
Schr$\ddot{o}$dinger equation (1D TDSE). However, the two models
only reproduce the experimental results qualitatively. The
calculations of the two models show large deviations in the momentum
range and the yield around zero momentum. The classical model in
\cite{5} discards the events for the returning electrons with
kinetic energy below the ionization potential of the singly charged
ion when recollision occurs. Therefore, the distribution around zero
momentum is excluded by this classical model. The 1D TDSE gives
distribution around zero momentum with yield much higher than the
experimental results because of the reduced dimensionality model.
Both models give a border momentum range than the experimental
results. For multi-cycles pulses, the ion momentum distribution is
symmetric since the amplitude of the electric field envelopes of
successive half-cycles are the same. While for few-cycle pulses,
they changes significantly and depend strongly on the
carrier-envelope phase (CEP). This gives rise to asymmetric ion
momentum distributions. For some CEP, there are two consecutive
half-cycles contributing effectively to NSDI and thus a doublet in
the ion momentum distribution may be observed.

Fig. 2 displays the corresponding correlated electron final-state
momentum distributions parallel to the laser polarization direction
for different CEPs, where the soft-core parameter $a=1.5$ a.u. for
Figs. 2(a)-2(e) and 1.0 a.u. for Figs. 2(f)-2(j) after first
ionization. White boxes indicate $\pm$2$\sqrt{U_p}$ =$\pm$1.66 a.u.
The correlated electron final-state momentum distributions have a
strong dependence on the CEP of the few-cycle pulse. The total
double ionization yields from the first and third quadrants are much
higher than those from the second and fourth quadrants. Furthermore,
the double ionization yields from the first and the third quadrants
depend strongly on the CEP, while that from the second and the
fourth quadrants depend slightly on the CEP.

Double ionization events due to (e, 2e) mechanism can only be found
in the first and the third quadrants of the final-state
electron-electron longitudinal momentum distributions, while due to
RESI mechanism can be found in all quadrants because of the
subsequently independent emission of the electrons [18]. Therefore,
the shape of final-state the electron-electron momentum
distributions in Fig.2 shows that both mechanisms are significant to
double ionization. Comparing Figs. 2(f)-2(j) with Figs. 2(a)-2(e)
respectively, we can find that double ionization yields from the
second and fourth quadrants increases a little when parameter a
decreases from 1.5 to 1 a.u. after first ionization. According to
\cite{24}, if the delay time between closest recollision and double
ionization is less than 0.25 laser cycle, the physical mechanism of
double ionization is classified to RCI and the mechanism is RESI if
the delay time larger than 0.25 laser cycle. The statistic results
show the double ionization yield due to RESI is about 50\% when
$a$=1.5 a.u. after first ionization. While the double ionization
yield due to RESI increases to above 60\% when $a=1$ a.u. after the
first ionization. The enhancement of contribution of RESI leads to
the ion yield around zero momentum increasing. This is well
understandable because the nuclear Coulomb attraction to the
electrons becomes larger as parameter a decreases. As a result, the
two electrons are hard to ionized through RCI mechanism. A lower
shielding parameter $a$ implies a more important role of the nuclear
Coulomb attraction in recollision. The good agreement between our
simulations and the experiments indicates that $a=1$ for Ar atoms is
the most suitable to describe the role of nucleus in recollision.
While $a>$1 or $a<$1 underestimates or overestimates the role of
nucleus in recollision respectively. This is why the ion momentum
distribution for $a=1.5$ and $\phi=1.3\pi$ is asymmetric.

\begin{figure}[htb]
\centering\includegraphics[width=12cm]{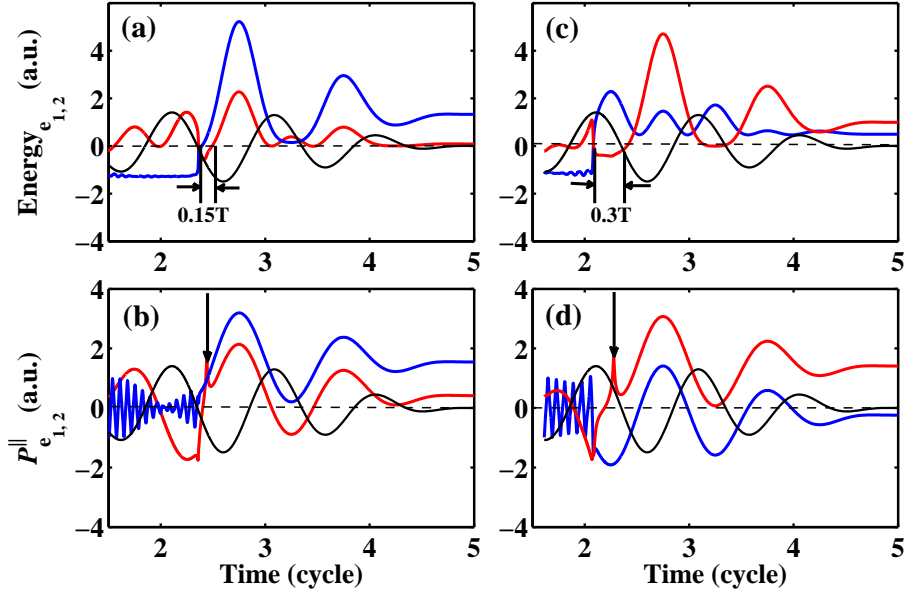} \caption{ (color
online).Typical trajectories for NSDI: (a) and (c) are energy
trajectories. (b) and (d) are momentum trajectories. Where the blue
and red curves indicate the first and the second electrons,
respectively. The black curve represents the electric field of the
laser pulse. (a) and (b) correspond to the electrons from the first
quadrant of Fig. 1(f). (c) and (d) correspond to the electrons from
the second or fourth quadrant of Fig. 1(f). }
\end{figure}

In order to explore the influence of nuclear Coulomb potential on
the electrons in double ionization process, we take advantage of the
back analysis of the classical model. Taking $\phi$=0.8$\pi$ as an
example, back analysis shows that there are two typical trajectories
displayed in Fig. 3. One of the typical trajectories [see Figs. 3(a)
and 3(b)] comes from the first quadrant of Fig. 2(f). Firstly, the
first electron is field-ionized (the red line) and thus possesses an
positive energy. It's energy and momentum is determined by the laser
electric field. The second electron is bounded by nucleus (the blue
line) and possesses an negative energy. At about 2.35 laser cycles,
the momentum of the first electron changes suddenly, decreasing to
zero rapidly. This implies that recollision occurs. At the same
time, the energy of the second electron rapidly increases and is
directly recollision-ionized in a very short time interval. The
first electron is firstly bounded and then ionized by the laser
electric field. Before ionized, the momentum of the first electron
changes suddenly [as indicated by the arrow in Fig. 3(b)], implying
that the nucleus imposes a strong attraction of this electron.
Finally, the dynamics of both electrons are determined by the laser
electric field. This process spends about 0.15 laser cycle. The
other typical trajectory [see Figs. 3(c) and 3(d)] comes from the
second quadrant of Fig. 2(f). The delay time between recollision and
double ionization lasts about 0.3 laser cycle. This trajectory shows
the similar behavior [indicated by the arrow in Fig. 3(d)]. The
trajectory analysis reveals that the nuclear Coulomb attraction
plays an important role in the microscopic dynamics in the process
between recollision and final double ionization.

The above results verifies that the soft-core parameter $a=1$ a.u
after first ionization is suitable to describe the role of nuclear
Coulomb attraction in recollision process of NSDI of Ar atoms. It is
confirmed once again by the simulation results of NSDI of Ar atoms
with laser intensities below the recollision threshold, which are
given below. Therefore, our theoretical simulations provide a
reliable benchmark for determining the CEP of the few-cycle pulses
by comparing with the experimental results using phase-dependent
NSDI. Fig. 4 shows the ratio of total NSDI yields from the first and
from the third quadrants as a function of the CEP. The ratio
increases rapidly for the CEP ranging from $0.2\pi$ to $0.7\pi$ and
decreases rapidly for the CEP ranging from $1.1\pi$ to $1.5\pi$. The
ratio has an analogy with that of total ATI yields of atoms in
few-cycle pulses from the left and the right sides [25].

\begin{figure}[htb]
\centering\includegraphics[width=5cm]{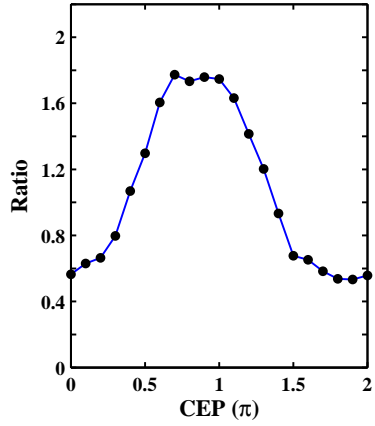} \caption{ (color
online). Ratio of the total NSDI yields from the first and the third
quadrants as a function of the CEP. The all calculated parameters
are the same as in Fig. 1. The circles are the calculated data. The
solid line is drawn to guide the eye. }
\end{figure}

\begin{figure}[htb]
\centering\includegraphics[width=10cm]{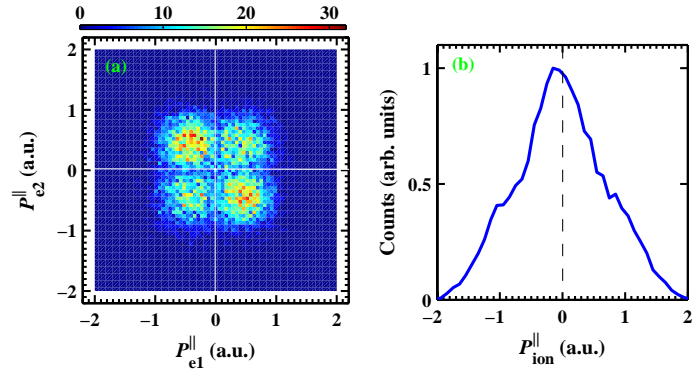} \caption{(color
online). (a) Correlated longitudinal momentum distribution
$P^{||}_{e1}$ vs $P^{||}_{e2}$ for argon double ionization. (b)
Longitudinal momentum spectra of Ar$^{2+}$ ions. The intensity
$I$=0.7$\times$10$^{14}$ W/cm$^2$.}
\end{figure}

Finally, we give the simulation results of NSDI of Ar atoms by an
intense laser pulse with intensity below the recollision threshold.
The pulse is trapezoidal shape with two-cycle turn on, six cycles at
full strength and two-cycle turn off. The wavelength $\lambda$=800
nm and the intensity $I$=0.7$\times$10$^{14}$ W/cm$^2$, are the same
as in the experiment of Ref. \cite{10}. Here, the employed ensemble
contains two million classical electron pairs. Figs 5(a) and 5(b)
show the correlated electron and doubly charged ion longitudinal
momentum distributions respectively for $a=1$ a.u. after first
ionization. Two prominent features are seen in Fig. 5(a): the
dominance of events for electron emission into opposite hemispheres
and a clear minimum at the origin. The ion longitudinal momentum
spectra in Fig. 5(b) shows a sharp maximum around zero momentum, in
contrast to the characteristic double-hump structure of NSDI with
intensities above the recollision threshold. The resulting
correlated momentum distributions in Figs. 5(a) and 5(b) are in
excellent agreement with the experimental results in Figs. 2(b) and
2(d) of Ref. [10], respectively. Below the recollision threshold,
such perfect agreement once again prove that the value of the
soft-core parameters $a$ selected in this paper accurately describe
the nuclear Coulomb attraction in the double ionization process of
Ar atoms.

We also analyze the double ionization process for the laser
intensity below recollision threshold. Back analysis reveals that
the RESI mechanism is the dominant process and most of the
trajectories only include one collision. For the trajectories from
the first and third quadrants of Fig. 5(a), one electron ionized
immediately after recollision while the other electron is excited by
recollision and then emits with laser-assisted ionization just
before the subsequent peak of the electric field. For the
trajectories from the second and fourth quadrants in Fig. 5(a), both
electrons are excited by recollision, and then they get ionized step
by step around various electric fields \cite{14}. For this
intensity, the electrons spend a significant time at the excited
states after recollision, where the nuclear Coulomb force is
effective. Thus the nuclear Coulomb attraction plays an important
role in the ionization process of argon at this laser intensity.

The parameter $a$ after first ionization for Ar atoms is much
different from that for He atoms. Ref. \cite{20} has shown that when
the soft-core parameters $a$ after first ionization is very small
($<<1$ a.u.), the simulation results using the 3D classical
ensembles are in good agreement with experimental results
\cite{18,19}. The remarkable difference between the values of $a$
for He and Ar atoms essentially originates from the different atomic
structures. He possesses only two electrons, while Ar possesses
eighteen electrons. Except the two outer active electrons, the inner
sixteen electrons of Ar exert a shielding effect on the nuclear
Coulomb attraction to the two outermost electrons. As a consequence,
it leads to suppressed influence of Ar nucleus in NSDI process,
compared with the case of He. This implies that the influence of the
inner electrons for many-electron atoms plays a significant role in
the dynamic details in strong-field double ionization process.

\section{Summary}
In summary, we have exploited the 3D classical ensembles to
investigate the strong-field NSDI of Ar atoms. By adjusting the
nuclear Coulomb potential to some proper form after first
ionization, we can quantitatively reproduce the experimental
correlated electron and doubly charged ion momentum spectra from
NSDI of Ar for both laser intensities above and below the
recollision threshold. This means that the 3D classical ensembles
can provide reliable simulation supports for determining the CEP of
few-cycle pulses using NSDI by comparing with experimental
measurements. Trajectory analysis reveals a sudden momentum change
of one electron, implying that the nuclear Coulomb attraction plays
a key role in the microscopic dynamics in process at and after
recollision. The significant difference of nuclear Coulomb
potentials for He and Ar implies the unignorable influence of the
inner electrons of Ar on NSDI.

\section*{ACKNOWLEDGMENTS}
This work was supported by the National Natural Science Foundation
of China under Grant No. 10774054, National Science Fund for
Distinguished Young Scholars under Grant No.60925021, and the 973
Program of China under Grant No. 2006CB806006.
\end{document}